\documentclass [12pt] {article}

\begin {document}


\noindent { \Large Nonlocal Astroparticles in Einstein's Universe}

\bigskip
\noindent  {I. Bulyzhenkov-Widicker} 

 
\bigskip
\begin {abstract}
{Gravitational probes should maintain spatial flatness for Einsten-Infeld-Hoffmann dynamics of re\-la\-tivistic matter-energy. The continuous elementary source/particle in Einstein's gravitational theory  is the $r^{-4}$ radial energy density rather than the delta-operator density in empty-space gravitation. The space energy integral of such an infinite (astro)particle is finite and determines its nonlocal gravitational charge for the energy-to-energy attraction of other nonlocal (astro)particles. The non-empty flat space of the undivided material Universe is charged continuously by the world energy density of the global ensemble of overlapping radial particles. Nonlocal gravitational/inertial energy-charges incorporate Machian relativism quantitatively into Einstein's gravitation for self-contained SR-GR dynamics without references on Newton's mass-to-mass attraction.  }
\end {abstract}
 \bigskip


The non-Newtonian geodetic precession of the asymmetrical Earth-Moon gyroscope in the gravitational field of the Sun was first noted by de Sitter\cite{Sit} in 1916. The modern quantitative analysis of the lunar-laser-ranging (LLR) data, accumulated by NASA between 1970 and 1986, was finally revealed\cite{Sha} in 1988. These data have been interpreted through the post-Newtonian parameter $\gamma$ related to supposed departure of 3D space from Euclidean flatness. Warping of empty space should unavoidably accompany point source models employed by Schwarzschild and Droste in the 1916 solutions\cite{Schw} to Einstein's 1915 equation. Nowadays everyone agrees that the energy-momentum density, rather than point mass, is a real source of General Relativity (GR) fields in this nonlinear gravitational equation for classical matter. From here Newtonian radial field energy density, $ \propto (\nabla U/m)^2 $,  works as a nonlocal $r^{-4}$ material particle in Einstein's gravitation. In other words, realistic space-time-energy organizations in Einstein's Universe cannot operate in principle with the empty (or free) space paradigm, because the GR energy-particle is distributed along with its radial field  over  non-empty space filled everywhere by particle's gravitational energy. 

Nonetheless, the numerical coincidence (2\% experimental error in 1988 and 0.7\% in 1996) of measurements with the Schwarzschild parameter $\gamma \equiv 1$ for the pre-quantum point particle was accepted by many researchers as an ultimate proof of space curvature that stroked out flat-space proponents from main gravitational conferences. Sommerfeld, Brillouin, Schwinger and many other classics had already explained the conceptual role of Euclidean space for modern physics.  Feynman even insisted on conference registrations under a pseudonym (for example in 1957, Chapel Hill, NC) in order to express his distain to unphysical interpretation of gravitational phenomena. In 1939 Einstein\cite{Ein} (and Narlikar\cite{Nar} in 1985) finally rejected the unrealistic Schwarzschild's solution, but the black hole  generation of modern cosmologists is persistently ignoring `irrelevant' criticism of their `reluctant father'.

First data of the Gravity Probe B mission\cite{Eve} could be also gladly accepted by the Schwarzschild-Schiff model with ${\bf \Omega}_G = (2^{-1}+\gamma){\bf v}\times {\bf \nabla} U/m $ for the geodetic precession frequency of four-vector spins without radial structures in locally curved space\cite{Sch}.  However, the superficial modeling of  (distributed) bodies through point spins in empty-space has prevented timely recognition of the GP-B finding - 3D space is strictly flat (better than 1\%) with respect to translations and rotations. The point is that available GP-B data (einstein.stanford.edu) for small spherical gyroscopes reiterate the same geodetic precession of the Moon-Earth gyroscope. This distributed system is well discussed without point singularities and its slow-motion precession is related in Einstein's relativity to inhomogeneous time dilation over the Moon orbit, rather than to local space curvature in question at the gyroscope center of inertia. 

Recall that Weyl completed his correct computations for non-point re\-la\-tivistic tops in 1923, well before the Einstein-Infeld-Hoffmann equation\cite{EIH}  of slow relativistic motion was derived in 1938. The similar post-Newtonian equations for slowly moving and rotating GR systems having finite dimensions and active/passive masses were also obtained by other relativists\cite{Edd}. And the classical Lagrange formalism for the Einstein-Infeld-Hoffmann dynamics very clearly specifies the enhanced geodetic precession of non-point orbiting gyroscopes through GR's time dilation or the $g_{oo}$ metric component\cite{Lan}. 

Why was the known time-dilation nature of the de Sitter - Weyl - Einstein - Infeld - Hoffmann geodetic precession never mentioned by contemporary researchers as the original GR alternative to very popular today curve-space interpretations of Einstein's physics? Einstein never refused from his 1938 post-Newtonian dynamics of distributed passive-active masses and tried to introduce non-point gravitational sources even for the 1915 covariant equation.  Initially Einstein and Grossmann put Newtonian potential $U/m = - GM/r$ only into the time subinterval in the Minkowski space-time interval\cite {Gro}. Lately Schwarzschild's point matter constructions reconnected both subintervals with the gravitational potential and badly redirected the Einstein-Grossmann metric project into empty-space frames of pre-quantum physics. 

Again, why were very strict anti-Schwarzschild statements of Einstein never cited by today's investigators of `Einstein's Universe'?  Unrealistic point particles and point spins are very useful sometimes for simplified, model interpretation of elementary matter-energy. But they should not substitute or strike out more rigorous pro-Einstein approaches to self-organization of space-time-energy in the nonlinear GR equations with nonlocal matter-energy distributions or with non-empty space.   All GR flowers should blossom and all solutions to the Einstein equation for empty and non-empty spaces should be equally discussed and respectfully compared by scientific forums. Professional reports cannot  ignore Einstein-Infeld-Hoffmann physics for slowly  rotating distributions of mass-energy in favor of Schwarzschild-Schiff mathematics for point particle-spins in question. What are the reasons to modify Einstein's physics prior to its tests? 

 LLR and GP-B data releases, for example, have firmly confirmed GR's time-dilation in the Einstein-Infeld-Hoffmann approach to distributed gyroscopes without any anti-Einstein contributions from spatial curvature in question. Therefore, there is no rational need at the moment to reinterpret the classical GR rotation through point-spin innovations. One can keep the 1923 Weyl-Einstein non-point gyroscope in order to compare the `Entwurf' flatspace generalization\cite{Gro} of the Minkowski interval with modern experimental data for spin-orbit and spin-spin frame-dragging (by assuming conservation of the total angular moment). Any mass-energy density may be conjugated only with local time, not with space coordinates. This conjugation can advise 
 non-empty, energy 
  space interpretation of Einstein's physics for the global overlap of $r^{-4}$ material carriers in the flat Universe of elementary energies.
 
The integration of particles into spatial structures of their fields was indeed assumed by Einstein: `We could regard matter as being made up of regions of space in which the field is extremely intense... There would be no room in this new physics for both field and matter, for the field would be the only reality,' translation\cite {Ton}. An extended particle has not yet been introduced analytically into the classical  theory of fields, which operates instead with the Dirac delta-density for the formal representation of  point particles in distant field points of empty space. 
The unreasonable formal assignment of different coordinate arguments to different parts of the analytical Poisson-Einstein equation for the `simplest' field of one point mass may be considered as our mathematical motivation to replace the point classical particle with the $r^{-4}$ radial energy-density  distribution.  Our physical motivation to look at the global spatial overlap of all material particles  originates from Newtonian gravitational stresses in an invisible material medium (called aether) connecting visible frames of interacting astronomic bodies.          
 
We rely on the metric formalism\cite{Gro} of General Relativity (GR) in our novel approach to Newton's aether and ancient Greeks' nonlocality of material {\it forms} extending 
visible frames of matter. 
  Let us consider a non-rotating probe particle with 
  passive  mass $m_p$ in a static central field of a gravitational source with the motionless active mass $M_a$ and active GR energy $E_a$. The passive GR energy ${\cal E}_p$ of  the probe particle, 
\begin{equation}
	{\cal E}_p \equiv {{m_pc^2 {\sqrt {g_{oo}}}  }\over {\sqrt {1-v^2c^{-2}}}} \equiv 
  {\cal K}   + {\cal U}_\Delta,
\end{equation}
incorporates its Special Relativity (SR) mechanical energy ${\cal K}$ and the so far undefined contribution of gravitational energy ${\cal U}_\Delta$. We are not going to employ Newtonian references from the classical mass-to-mass gravitation for this potential energy, ${\cal U}_\Delta = {\cal U}_\Delta (r)$, which is an unknown radial function of distance $r$ between centers of active, $E_a$, and passive, ${\cal E}_p$, nonlocal energies. We plan to use only SR energy references ${\cal K} \equiv m_pc^2 /{\sqrt {1 - v^2c^{-2}}} $ for gravitating mass-energies. Then, Einstein's SR and GR may be discussed together as the self-contained SR-GR theory for energy-to-energy interactions of non-local inertial energies or GR charges.

Based on `new' SR references for the mechanical energy ${K}$ in (1), one can redesign the GR metric component ${{g_{oo}}}$ of the pseudo-Riemannian metric tensor $g_{\mu\nu}$ in the following way,
\begin{equation}
	{\sqrt {g_{oo}}}\equiv {{ {\cal K} {\sqrt {1-v^2c^{-2}}}  }\over m_p c^2} + {{{\cal U}_\Delta {\sqrt {g_{oo}}}  }\over  { \cal E}_p} 	\equiv {{ {\cal K} {\sqrt {1-v^2c^{-2} }   }\over m_p c^2 (1 -{\cal U}_\Delta{\cal E}_p^{-1} ) }} \equiv {{ 1 \over 1 -{\cal U}_\Delta{\cal E}_p^{-1}  }}.
	\end{equation}
	
This metric component equally contributes to local physical time $d\tau  = {\sqrt {g_{oo}}}dt$ for an  imaginary point observer and to local proper time for probe mass centers in a static field without rotation (when $g_{oi} = 0$ and  $ds^2$ = $g_{oo}dt^2$  - $dl^2$). The active mass $M_a$ possesses a distributed active energy ${E}_a$ of the nonlocal GR source, while the passive mass $M_p$ can be associated with the distributed passive-inertial energy $E_p$ of the same radial carrier of matter, which have equal active and passive mass-energies. A spatial density of active energy of the distributed source-particle is locally balanced by passive sink-particle energy density in the vanishing\cite{Bul} Einstein curvature, $G_o^\mu \equiv g^{\mu\nu}R_{\nu o}  - 2^{-1}\delta^\mu_o g^{\rho\nu}R_{\rho\nu} = 0$, for every energy-momentum carrier with locally bound source-particle (active field) and  sink-particle (passive field) fractions of matter-energy.
Einstein maintained that all terms of his 1915 equation should be considered at field points and we  reiterated the continuous, field distribution of gravitational bodies within their spatial field structures by assigning active mass-energy to distributed source-fields and passive mass-energy to distributed sink-fields or distributed inertial particles. The 1907 Principle of Equivalence corresponds to the strict balance of active and passive energy-momentum components of every gesamt (=whole) energy carrier. Therefore, the Hilbert invariant conservation (1915) of energy-momentum densities for carriers with the zero balanced Einstein curvature $G_o^\mu = 0$ (for zero-temperature matter without the non-metric lambda-term   $\delta^\mu_o\Lambda$ or dissipative heat in $G_o^\mu + \delta^\mu_o = 0$)  generates equalities, 
   $G_{o;\mu}^\mu \equiv 0$ and $(u^\mu u_\nu G_o^\nu)_{;\mu} \equiv 0$, in full agreement with Klein's physical argumentations (1916), the  second Noether theorem (1917), and the differential (contracted) Bianchi identity.

   The GR non-empty space equation for the global summary of all local energy densities, $\sum_1^\infty u_\mu u_\nu G_o^\nu$ = 0, is valid for the world overlap of moving continuous sources and sinks, {\it i.e.} for the  joint nonlocal dynamics of all elementary carriers of active and passive integral four-momentums $m_{a/p}u_\mu$.  The Ricci-Tolman  mass-energy density $R_o^o$ of every gesamt carrier of paired active-passive (source-sink, yang-ying) energies is jointly contributed by equal densities of the distributed active source-energy $E_a$ and the distributed passive sink-energy $E_p$. Two non-vanishing affine connections, $\Gamma^o_{io} = \partial_i g_{oo}/2g_{oo}, 
\Gamma^i_{oo} $ = $ \partial_ig_{oo}/2$,  with one logarithmic potential $W \equiv - c^2 ln (1/ {\sqrt {g_{oo}}}) $ are  responsible for the Ricci scalar density,  
\begin {equation} 
\frac {R}{2} = R_o^o = g^{oo}R_{oo} =   g^{oo} \partial_i \Gamma^i_{oo} - g^{oo}\Gamma^i_{oo} \Gamma^o_{io} = \nabla^2 Wc^{-2} + (\nabla Wc^{-2})^2,
\end {equation}
of elementary material spaces with strong static fields in their rest frames, where $g_{oi} = g^{oi} = 0$,  $g^{oo} = 1/g_{oo}$, and $g_{ij} = -\delta_{ij}$. 
 Many relativists tend to drop $(\nabla W/c^2)^2$ next to the `linear' term $\nabla^2 W/c^2$ for Newtonian weak field limit ($- W/c^2 \approx - {\cal U}_\Delta/{\cal E}_p = + const/r << 1$) of the Ricci curvature $R^o_o$ in the Einstein Equation. 
  Such an erroneous approach to the Ricci tensor forma\-lism contradicts the Principle of Equivalence which requires local identities, $\nabla^2 W/c^2 \equiv (\nabla W/c^2)^2 $, of active and passive mass-energies for both strong and weak  gravitational interactions. Even without references on  Einstein's physics,  it is not reasonable mathematics when one claims for the week field limit ($W \approx - const/r$) that $\nabla^2 W \equiv r^{-1}\partial_r^2  (rW) \approx - r^{-1}\partial_r^2  const \equiv 0$ is the largest term in $R^o_o$. In fact, the 1915 Einstein equation $G_\nu^\mu = \kappa T_\nu^\mu$ cannot result mathematically in Newtonian gravitation, unless one 'simplifies' this overloaded tensor equation by geometrization of particles in $T^\mu_o$ through the Ricci tensor densities within the `vectorized' four-equation $G^\mu_o = 0$.

    Contrary to the qualitative interpretation of the Principle of Equivalence for empty space gravitation of point particles, non-empty space physics with Newtonian aether of nonlocal gravitational bodies can describe the local  equivalence of active and passive (inertial) mass-energy densities quantitatively, 
   \begin {eqnarray}
   \rho_a c^2 \equiv  \frac {c^2 \nabla^2 W} {4\pi G}  \equiv - {{c^4}\over 4\pi G r^2} \partial_r \left [r^2 \partial_r ln \left (\frac {1}{{\sqrt {g_{oo} }}}\right )\right ] \\ \nonumber
    = {{c^4}\over 4\pi G } {\left [\partial_r ln \left(\frac {1}{{\sqrt {g_{oo}}}} \right )\right ]^2} \equiv \frac {c^2 (\nabla W)^2} {4\pi G} \equiv \rho_p c^2, 
\end{eqnarray}
with peculiarity-free solutions for the metric tensor component (2) even in strong fields. General radial solutions of the nonlinear Poisson  equation (4),  $1-{\cal U}_\Delta{\cal E}^{-1}_p \equiv 1/ {\sqrt {g_{oo}}} = C_1 r^{-1} + C_2$, depends on two constants $C_1$ and $C_2$. One constant can be defined ($C_2 = 1$) due to the SR asymptotic behavior of the GR metric, $g_{oo}(\infty) \rightarrow 1 $. The other constant ($C_1 = GE_a/c^4$) can be found after the volume integration of the active  energy density from (4), 
\begin{equation}
	{ E}_a = \int_o^\infty \rho_a(r) c^2 4\pi r^2 dr = - {{c^4 r^2}\over G } \partial_r ln (1/ {\sqrt {g_{oo}}})|_{r\rightarrow o}^{r\rightarrow \infty}  =   {{{ c^4r^2} \partial_r ({\cal U}_\Delta {\cal E}_p^{-1})}\over G (1 - {\cal U}_\Delta{\cal E}_p^{-1})  } |_{r\rightarrow o}^{r\rightarrow \infty} . 
\end{equation}
The radial potential ${\cal U}_\Delta{\cal E}^{-1}_p = - C_1r^{-1}$ of the active nonlocal energy-charge $E_a = c^4 C_1/G $ for the passive (probe) nonlocal energy-charge ${\cal E}_p$ 
corresponds to the energy-to-energy attraction law\cite {Bul} \begin{equation}
	{{\cal U}_\Delta}  = - {{G E_a }\over  c^4 r}{\cal E}_p,
	\end{equation}
	in self-contained GR with SR references and Newtonian aether (4).
	
Again, we specified two constants $C_1 = GE_a/c^4 \equiv r_o$ and $C_2 = 1$ through the Principle of Equivalence for active and passive components of the Ricci-Tolman mass-energy in (3)-(4) and through the SR asymptotic metric. Therefore, we specified the GR metric tensor component $g_{oo} = ( 1 + r_or^{-1})^{-2}$ 
and the static flatspace metric $ds^2 = g_{oo}dt^2 - \delta_{ij}dx^id^j$  (for rest-frame fields without rotation or net spin) 
without references on Newtonian gravitation. The GR vector force in static central fields\cite {Lan},
 \begin {equation}
{\bf f} \equiv \frac {m_p }{{\sqrt {1 - v^2}}} \nabla ln \left ( \frac {1}{{\sqrt {g_{oo}} }}\right)   
= \frac {m_p {\sqrt {g_{oo}} } }{{\sqrt {1 - v^2}}} \nabla \left ( \frac {1}{{\sqrt {g_{oo}}}}\right) = - \frac 
{{\hat {\bf r}} GE_a } {c^4r^2} {\cal E}_p ,   
\end {equation}
is exerted upon the passive energy-charge ${\cal E}_p$, which is the measure of inertia for the 
probe mass $m_p$.
The strong field intensity ${\bf f}/{\cal E}_p$ keeps 
 Gaussian surface flux ($E_a = const$) for paired energy-to-energy interactions in flat space. 
	However, active/passive energy-charges are constant in (7) only in the absence of third bodies which can vary such gravitational/inertial charges in full agreement with Mach's  ideas\cite{Mac} embedded into the self-contained SR-GR dynamics (1)-(7) of nonlocal continuous energies or nonlocal gravitational bodies.

Based on the SR form of the mechanical part of the GR passive (probe)  energy ${\cal E}_p$, Einstein's metric theory quantitatively incorporates Machian ideas and rigorously relates the local component ${{g_{oo}}} = [1 + ( r_o/ r)]^{-2}$  to the nonlocal active energy $E_a$.  It is worth noting that the dynamical gravitational equation $\sum_{k=_1}^{\infty} u_k^\mu u_k^\nu (G^k_{o\nu} + g^k_{o\nu}\Lambda^k)= 0$ for all overlapping elementary energy carriers (nonlocal GR energies paired with their internal, non-metric energies $\Lambda^k$) maintains  that every  material carrier continuously occupies  the entire Universe despite the ultrashort gravitational scale $r_o=G E_a/c^4$ for the elementary matter density. Indeed, the active and passive mass-energy densities  of the radial astrocarrier with equal active/passive energy charges $E_{a/p} = r_o c^4/G = E$, 
 \begin{equation}
	\rho_a(r) c^2 \equiv \rho_p(r)c^2  =  E{{ r_o} \over 4\pi r^2(r_o + r)^2} 
	=  \frac {c^4}{4\pi G r^2 } {{1} \over [1 + (rc^4/GE)]^2}, 
\end{equation} 
exist everywhere, at all radial distances in the nonlocal elementary microcosm of each rest-mass formation.  Electrically bound elementary astrocarriers of active and passive radial GR energies constitute nonlocal molecules, nonlocal mechanical bodies (cosmism of all people), nonlocal planets, etc...  Ultrashort microscopic  scales, $r_o = GE_{atom}/c^4$, of electrically neutral atoms on bodies' visible surfaces are beyond  the Planck quantum length, instrumental resolutions, and the perception level. Nonetheless, all surface and bulk atoms of observed gravitating bodies are nonlocal astrodistributions of radial active plus passive energies, while centers of spherical symmetries of these atoms belong to the most dense (visible) frames of an infinite body (with its low-dense invisible aether). Continuous fields and continuous particles in nonlocal energy-to-energy gravitation
 are (yin-yang) paired distributions of equal amounts of matter-energy.


   Flat material space with Newtonian aether, specified by the astrodistribution (8) with the radial particle  density $n(r)=r_o/4\pi r^2(r + r_o)^2$, differs in principle from Schwarz\-schild's `point matter - empty space' model of physical reality (roughly described by  operator mathematics of the non-analytical density $\delta (r)$ for the point particle approximation).  Therefore, the Birkhoff theorem for empty (but curved) 3D spaces cannot be relevant to our reading  of  Einstein's physics with joint geometrization of particles and fields in non-empty (but flat) space. Moreover, our static metric solution $ ds^2 = dt^2\left (1 + {{r_o } r ^{-1}} \right)^{-2} - \delta_{ij}dx^idx^j$  for the universal attraction (6)-(7) of nonlocal passive  energy-charges by the nonlocal active (source) energy-charge has been derived to criticize {\it ad hoc} dogmas of the empty space world with point mass-energy peculiarities. Both elementary electric charges (with the Gauss flux conservation due to spatial flatness) and  masses co-exist for observations in one 3D space, which therefore must have only one universal, common 3D sub-geometry for all kinds of particles or fields. Otherwise, how warped 3D interval for electron's mass might allow it to move together with electron's Gauss flux center without  spatial splits under non-warped Euclidean translations for electricity?


\end {document}